\documentstyle[preprint,aps]{revtex}   
\begin{document}
\draft

\title{Dynamics of Random  Networks: Connectivity and
First Order Phase Transitions}

\author{ Albert-L\'aszl\'o Barab\'asi} 
\address{Department of
Physics, University of Notre Dame, Notre Dame, IN 46556.}

\date{\today}

\maketitle

\begin{abstract}
The connectivity of individual neurons of large neural networks
determine both the steady state activity of the network and its answer
to external stimulus.  Highly diluted random networks have zero
activity. We show that increasing the network connectivity the
activity changes discontinuously from zero to a finite value as a
critical value in the connectivity is reached.  Theoretical arguments
and extensive numerical simulations indicate that the origin of this
discontinuity in the activity of random networks is a first order
phase transition from an inactive to an active state as the
connectivity of the network is increased.
\end{abstract}

\pacs{}



Networks of neuron type threshold elements have generated a lot of
interest lately, motivated by their potential for reproducing
neurobiological processes and understanding the generic mechanism
governing basic brain functions.  Most of the studied models deal
either with fully connected networks (for which every neuron is
connected to all neurons in the system \cite{Amit}) or highly diluted
networks \cite{Derrida87,Kurten88}.  At this point little is known
about the properties of networks with arbitrary connectivity,
including networks whose connectivity is realistic from the standpoint
of biology.

Here we take a first step towards understanding the dynamical
properties of random networks by investigating the effect of
connectivity on the network dynamics \cite{egyeb}, where the
connectivity is the probability that two randomly selected neurons
have a synaptic connection.  A network with small connectivity is
inactive, i.e. if we excite it with an external stimulus, the activity
dies out in a short time. One would expect that increasing the
connectivity leads to a gradual increase in the activity of the
network.  Contrary to this simple picture, here we show that upon
increasing the connectivity, the steady state activity does not
increase continuously, but jumps discontinuously, the network going
through a {\it first order phase transition} from an inactive to an
active state.  This result may help us understand the evolutionary
driving forces that lead to the synaptic densities characterizing the
brain by providing the range of connectivities necessary for the
network to display a nonzero activity, and sheds light into the
difficulties occurring during modeling of low level sustained activity
in the cerebral cortex \cite{abeles}.

Consider a network formed by $N$ neurons.  Each neuron is connected
randomly to any of the other neurons with a probability $\rho$,
i.e. $\rho$ is measure of the network connectivity.  The detailed
wiring diagram of the system is given by the matrix ${\bf J}$, whose
element $J_{ij}$ is 1, if there is a directed path from neuron $i$ to
$j$, and zero otherwise.

First we consider a simplified version of the McCulloch-Pitts model
\cite{Pitts}, limiting the model to only excitatory neurons.  Later we
will show that this simplification does not affect the main
conclusions of the paper.  Every neuron $i$ can be in two possible
states, active ($s_i=1$) or inactive ($s_i=0$).  Using synchronous
updating, the state of neuron $i$ at time $t+1$ is $s_i(t+1)=1$ with
probability $P(h_i, T)= [1+\tanh(h_i(t)/T)]/2$, and $s_i(t+1)=0$
otherwise. Here $h_i(t) =\sum_j J_{ji} s_j(t)-\theta$, $\theta$ is the
threshold of the individual neuron, and   $T$ (temperature)
characterizes the spontaneous firing of a neuron.  The {\it activity}
of the network can be characterized by the normalized number of active
neurons at time $t$: $x(t)=(1/N)\sum_i s_i(t)$.  Hereafter we call a
network {\it active} ({\it inactive}) if $x(t) \neq 0$ ($x(t) = 0$),
for large $t$.

After activating a fraction $x$ of all neurons, if the network is
highly diluted ($\rho \to 0$), the neurons are not connected to
sufficient number of other neurons to overcome $\theta$, and the
activity decays to zero. In contrast, a highly connected network
($\rho \to 1$) for almost any initial conditions leads to an
active
state. The steady state dynamics in the active state is rather
complex, leading to periodic oscillations at $T=0$, whose period
depends
on
$N$, $\rho$ and ${\bf J}$.  However, at  nonzero  temperatures
the
dynamics is quasiperiodic, i.e.  the system explores randomly a
number
of periodic orbits.

Our goal is to understand the changes in the network activity as the
connectivity $\rho$ is varied. For this we implement a numerical
method to calculate the {\it free energy}, that characterizes the
dynamics of networks with fixed ($N, \rho$).  For now we limit
ourselves to symmetric networks, for which $J_{ij}=J_{ji}$.

Consider the above network model after it reached its steady state and
define $f'(x)$ as \begin{equation} x(t+1) \equiv x(t)+f'(x).
\label{flow-disc}
\end{equation}
Here $f'(x)$
depends on ${\bf J}$ and on the initial conditions
$\{x_i(t=0)\}$.
Averaging over the various realizations of the network topology,
${\bf
J}$,  leads to $f(x)=[f'(x)]_{J,x(0)}$,
which is a smooth and univalued function.  Alternatively,  we can
define
the continuum version of (\ref{flow-disc}), using the averaged
$f(x)$
\begin{equation}
\partial_t x(t) = f(x(t)) + \eta(t),
\label{flow-cont}
\end {equation}
where we added the  uncorrelated noise,  $\eta(t)$,  to
incorporate the randomness
of the updating rule at nonzero  temperatures. 

Formally,  (\ref{flow-cont}) can be generated from the {\it
free energy}
${\cal F}(x)$ using
\begin{equation}
\partial_t x(t)= -  {\partial {\cal F}(x) \over \partial x} +
\eta(t),
\end{equation}
where
\begin{equation}
{\cal F}(x) = - \int_0^x f(x') dx'.
\label{def-free}
\end{equation}
As we show below, using (\ref{flow-disc}) and (\ref{def-free}), ${\cal
F}(x)$ can be {\it calculated} using a mean field approximation and
{\it measured} in numerical simulations.

{\it Mean field theory (MFT)---} At time $t$ we randomly activate a
fraction $x(t)$ of neurons. The goal is to calculate $x(t+1)$. The
calculation proceeds in three steps.

(i) Choosing randomly a neuron,  the probability $P(N,k, \rho)$
that it is  connected to  $k$ other neurons
  follows the  binomial distribution
$P(N,k,\rho)=  {N-1 \choose k}  \rho^k
(1-\rho)^{N-1-k}$,
where ${ a \choose b}= a!/b!(a-b)!$.

 (ii) Knowing that a fraction $x$ of all neurons are active, the
probability $\Pi(k,m,x)$, that of the $k$ neurons the chosen
neuron
is connected to, $m$ are active, is given by the binomial
distribution
$\Pi(k,m,x) = {k \choose  m}  x^m
(1-x)^{k-m}. $
 (iii) Finally, a neuron with $m$ active neighbors is activated
with
probability $P(m,T)$.

Thus, the activity of the network, $x(t+1)$, is given by
\begin{equation} x(t+1)= \sum_{m=0}^{N-1} P (m,T)
\sum_{k=m}^{N-1} P(N,k,\rho)\Pi(k,m,x(t)).  \label{xt}
\end{equation}
This expression provides $f(x) \equiv x(t+1)-x(t)$, and using
(\ref{def-free}), we can calculate the free energy ${\cal F}(x)$.

Obtaining (\ref{xt}) we made two approximations.  In (i) we
neglected
the quenched nature of the randomness in the connectivity matrix
${\bf
J}$, replacing ${\bf J}$ with a new one at every time step,
subject to
fixed $\rho$ and $N$.  In (ii) we assumed that every neuron has
the
same probability $x$ to be active, independent of the network
topology. In reality the activity of a neuron is highly
correlated
with its connectivity. However, even within these
approximations, the MFT captures correctly the nature of the
phase
transition from the inactive to the active state.

{\it Phase transition---} (a) The free energies obtained from
(\ref{xt}) for $T=0$, $N=100$, $\theta=2$, and various values of the
connectivity $\rho$ are shown in Fig. \ref{fig1}(a). For small $\rho$
the only stable state has zero activity ($x=0$), i.e.  starting with
any $x$, the activity decays.  Increasing $\rho$, at $\rho_1(0)
\approx 0.042$ the free energy develops an inflection point, leading
to a second, local and unstable minima at $x_1(0)$, corresponding to a
nonzero activity.  Further increasing $\rho$, at $\rho_c(0) \approx
0.046$, ${\cal F}(x_1)$ becomes smaller than ${\cal F}(x=0)$,
indicating that the system undergoes a {\it first order phase
transition} during which the activity jumps from $x=0$ to $x_1 \neq
0$.  The formally stable $x=0$ state becomes a metastable state.

(b) For nonzero $T$, thermal fluctuations induce a thermal activity,
$x_0(T)=[1+\tanh(- \theta/T)]/2$, that is independent of the network
topology.  Expanding (\ref{xt}) for small $\rho$ and $x$, we find that
the derivative of the free energy in $x=0$ is negative for nonzero
$T$, thus ${\cal F}$ has no minima at $x=0$, but only at the thermally
activated $x_0(T)$.  For small $T$, increasing $\rho$ leads to the
appearance of the metastable $x_1(T) > x_0(T)$, which becomes a global
minima at $\rho_c(T)$ (see Fig.  \ref{fig1}(b)). Thus for small $T$ we
observe a first order phase transition from the thermally activated
state, $x_0(T)$, to the active state $x_1(T)$.

(c) For high $T$ the thermally induced activity, $x_0(T)$, dominates
the dynamics of the system, such that for $T \geq T_c$ the local
minima corresponding to the active state $x_1(T)$ does not appear, and
we can not distinguish between the thermally activated state and the
active state of the network (see Fig. \ref{fig1}(c)).

We summarize the above behavior using the $(\rho,T)$ phase diagram,
shown in Fig. \ref{fig1}(d). For small $T$ the active and the
thermally activated states are separated by the line $\rho_c(T)$.
However, the $\rho_c(T)$ line ends at $T_c$, and for $T > T_c$ the
system is too noisy to distinguish the thermally activated state from
the true activity of the network.

{\it Steady state behavior---} Due to the discussed approximations,
the MFT does not provide us the steady state behavior of the network.
For this we measured the steady state $f(x)$  using the discussed
network model.  Fig. \ref{fig2} shows the measured free energy for
$T=0$ and for $T > T_c$. For $T=0$ we observe a first order phase
transition from the inactive to the active states, as predicted by the
MFT (see Fig. \ref{fig2}(a)).  For $T \neq 0$, thermal fluctuations
lead to a thermally induced metastable state \cite{finite1}, which is
indistinguishable from the active state of the network if $T > T_c$.
Thus the phase diagram for the steady state behavior is similar to the
one predicted by the MFT, shown in Fig. \ref{fig1}(d).

Both the MFT and simulations in the steady state predict the
existence
of a first order phase transition for small temperatures as the
connectivity of the network is increased.  A number of important
questions arise at this point \cite{finite}: How generic is the
observed first order phase transition?  What are the implications
of
this transition?

To answer the first question, we studied extensions of the described
model, including elements that are often considered in neural network
simulations.  Simulations show that introducing asymmetry ($J_{ij}
\neq J_{ji}$) modifies the $\rho_c(T)$ curve, but the system still
undergoes a first order phase transition.  We find that the phase
transition is not affected by the inclusion of inhibitory neurons
either, nor by the nature of updating process (random or synchronous).
Thus we conclude that the existence of a first order phase transition
for small temperatures is a generic property of random networks, and
it is related to the topology of the network rather than the dynamics
of the individual neurons.

 What determines the actual connectivity and the synaptic density of
certain parts of the brain? Naturally, a complete answer to this
question should consider the genetically determined non-randomness in
the synaptic connections. However, our results suggest that for random
networks the optimal connectivity depends on the dynamical properties
desired for the particular network.  When a certain fraction of
neurons are activated by an external stimulus, the network should
respond with nonzero activity. For most brain functions this activity
should decay after some time if the stimulus is not sustained.  For a
large network this can be achieved by exciting the network into a
metastable state. In this case the activity decays after some
characteristic time, which depends on the free energy barrier between
the stable and the metastable states.  On the other hand, if sustained
activity is the goal, one needs to use a $\rho$ larger than
$\rho_c(T)$.  For a long time cerebral cortex modeling was halted by
the inability of large random network models to reproduce the observed
low level activities \cite{abeles}.  The observed first order phase
transition explains the origins of this failure: the arbitrary small
activity between zero and $x(\rho_c(T))$ is simply not available in
network models due to the jump in the activity at the transition
point.

Furthermore, the outlined method (\ref{flow-disc}-\ref{def-free})
could be used in exploring the properties of large networks when
only
a small fraction of the neurons can be monitored in the
laboratory. Examples include  multi-electrode recording
techniques,
that provide information about the activity of typically tens of
neural cells \cite{rem}.  Such measurements may help understand
the
network activity by providing ${\cal F}(x)$ for small $x$.

\begin{figure}
\caption{The free energy predicted by the mean-field theory
(\protect\ref{xt}).
(a) $T=0$. The curves, from top to bottom, correspond to
$\rho=0.020,
0.042, 0.046, 0.050$.  (b) $T=1.11$. The curves, from top to
bottom,
correspond to $\rho=0.020, 0.042, 0.043, 0.045$.  (c) $T=10.$ The
curves, from top to bottom, correspond to $\rho=0, 0.020, 0.070,
0.101$. (d) The $(T, \rho)$ phase diagram. The solid line
corresponds
to $\rho_c(T)$, separating the thermally activated phase from the
active state. The error bars are smaller than the symbols. }
\label{fig1}
\end{figure}

\begin{figure}
\caption{The steady state free energy obtained from numerical
simulations for $N=100$ and $\theta=2$. Average over 1000
runs is taken.
(a) $T=0$. The curves, from top to bottom correspond to
$\rho=0.005,
0.010, 0.011, 0.015$.  (b) $T=10.0$. The curves, from top to
bottom,
correspond to $\rho=0.0, 0.030, 0.043, 0.101$. }
\label{fig2}
\end{figure}



\begin{references}
\bibitem{finite1} The thermally activated minima is difficult to
visualize using the resolution determined by the system sizes
used in
the simulations.
\bibitem{finite} To fully exploit the presented results we need to
understand how the parameters characterizing the first order phase
transition depend on the size of the network.  Preliminary results
indicate that in the MFT, for $T=0$, $\rho_c(T,N) \sim 1/N$.
\end{references}
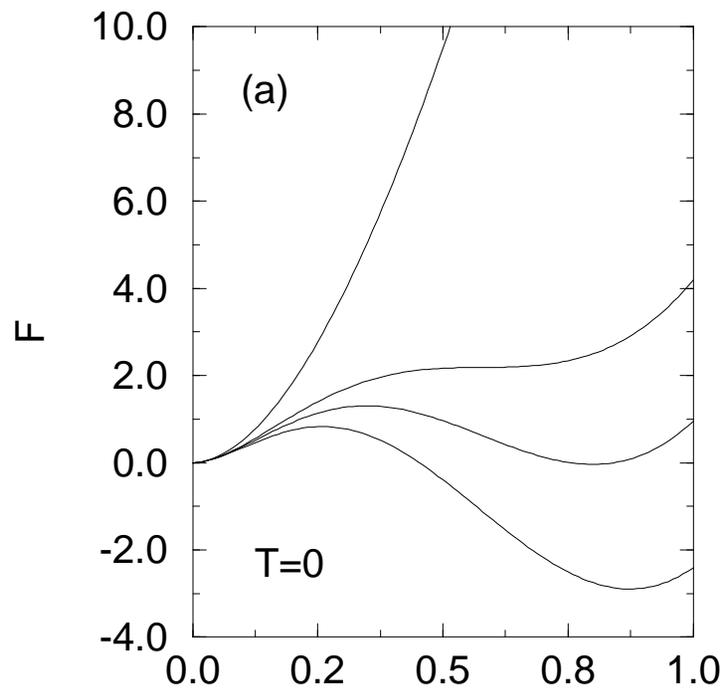
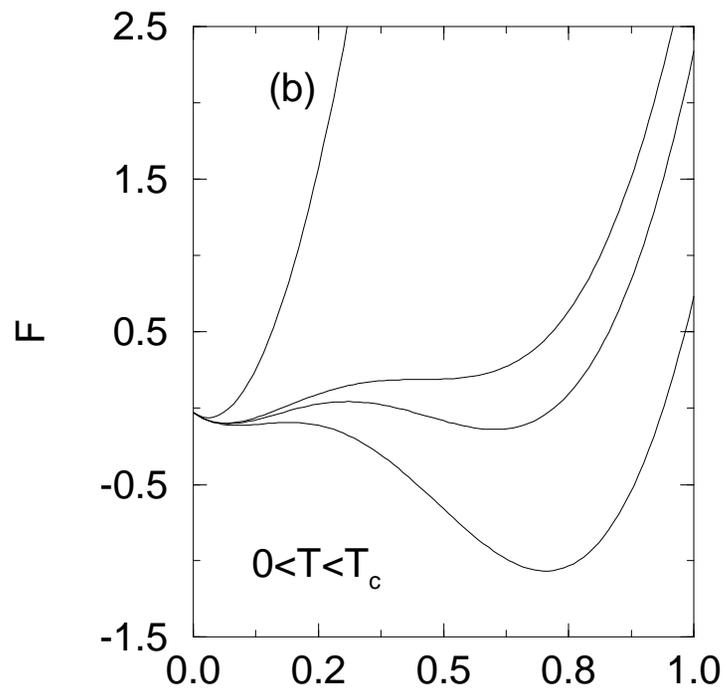
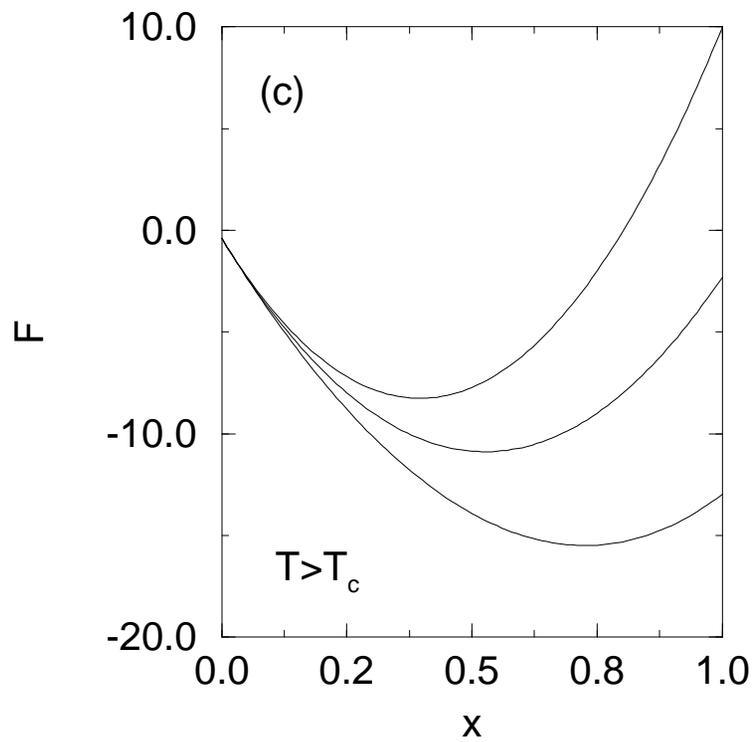
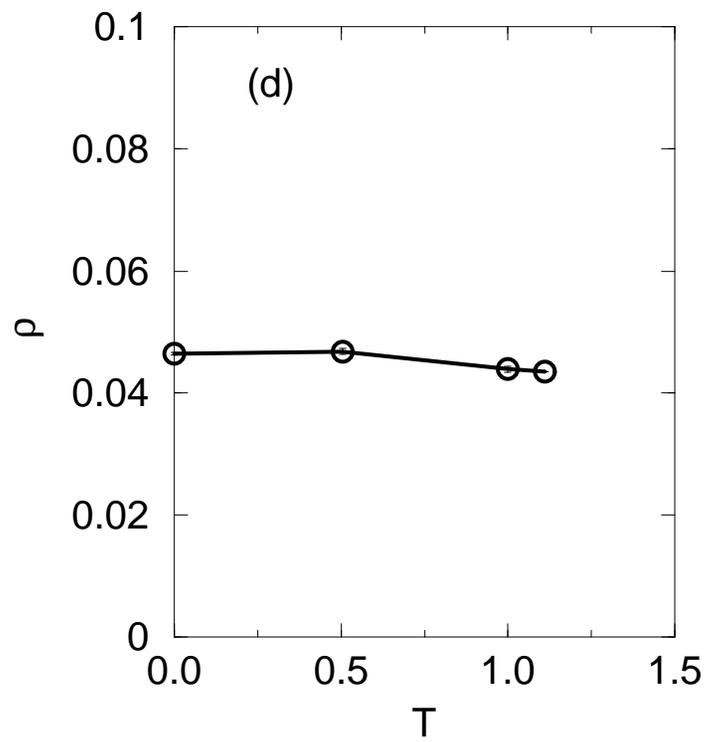
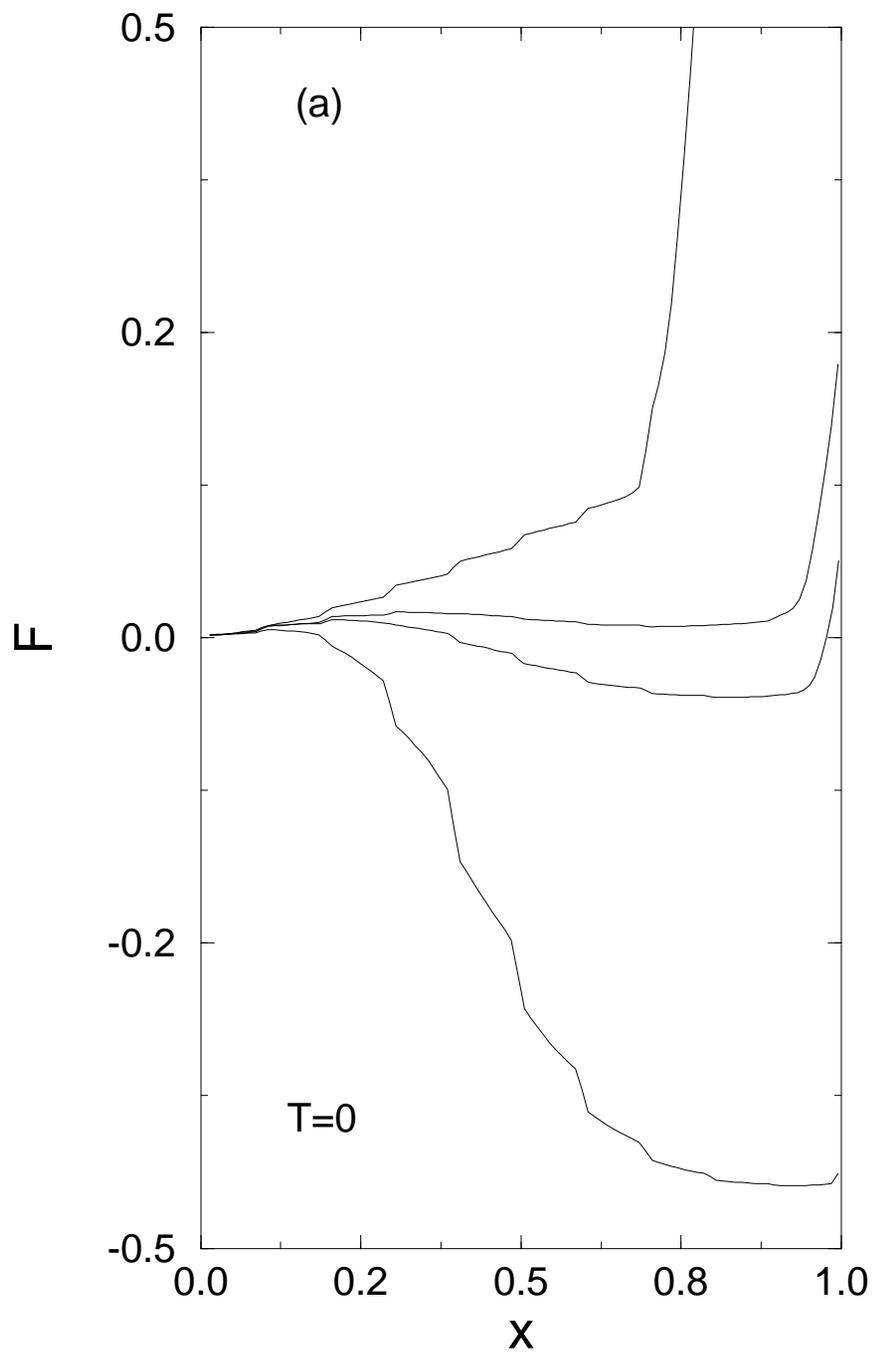
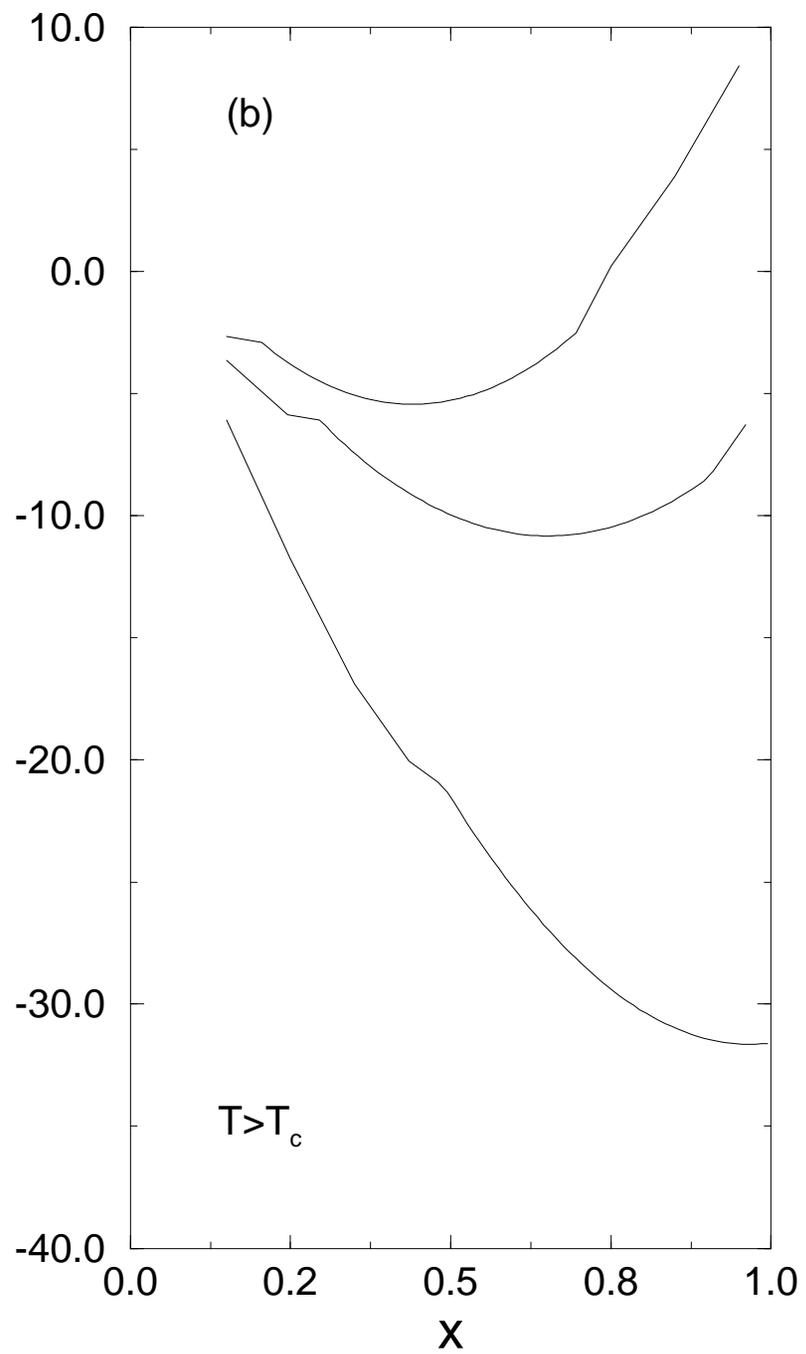
\end{document}